%% file: main.tex
\documentclass[journal,12pt,onecolumn,draftclsnofoot,]{IEEEtran}
\usepackage[utf8]{inputenc}
\usepackage{cite}
\usepackage{amsmath,amssymb,amsfonts}
\usepackage{algorithmic}
\usepackage{nicematrix,tikz}
\usepackage{graphicx}
\usepackage{textcomp}
\usepackage{array,booktabs,makecell}
\usepackage[font=footnotesize]{caption}
\usepackage{float}
\usepackage{tikz}
\usepackage{cite,balance}
\usetikzlibrary{shapes.geometric, arrows}
\usepackage{pgfplots}
\usepackage{caption}
\usepackage{subcaption}
\newcommand\numberthis{\addtocounter{equation}{1}\tag{\theequation}}
\usepackage{algorithm}
\usepackage{xcolor}
\definecolor{OliveGreen}{rgb}{0,0.6,0}
\usepackage{algorithmic}
\def\BibTeX{{\rm B\keri-.05em{\sc i\keri-.025em b}\keri-.08em
    T\keri-.1667em\lower.7ex\hbox{E}\keri-.125emX}}
\usepackage{multirow}

\usepackage{amsthm}
\newtheorem{remark}{Remark}
\newtheorem{prop}{Proposition}

\author{Shudi Weng, Fan Jiang, and Henk Wymeersch}
\date{}
\title{Wideband mmWave Massive MIMO Channel Estimation and Localization
}

\begin{document}

\maketitle

\begin{abstract}
Spatial wideband effects are known to affect 
channel estimation and localization performance 
in millimeter wave (mmWave) massive multiple-input multiple-output (MIMO) systems. Based on perturbation analysis, we show that the spatial wideband effect is in fact more pronounced than previously thought and significantly degrades performance, even at moderate bandwidths, if it is not properly considered in the algorithm design. We propose a novel channel estimation method based on multidimensional ESPRIT per subcarrier, combined with unsupervised learning for pairing across subcarriers, which shows significant performance gain over existing schemes under wideband conditions. 
\end{abstract}


\section{Introduction}
The integration of the localization and communication systems has attracted increasing attentions in 5G and beyond \cite{HexaX21, gupta2020sdnfv, chen2022tutorial}. In particular, massive multiple-input multiple-output (MIMO) and millimeter wave (mmWave) techniques have been demonstrated to significantly improve the communication performance in terms of reliability, throughput, and scalability. At the same time, the deployment of large antenna arrays and the availability of large bandwidth over mmWave frequencies improve the angle and delay resolution, leading to high accuracy localization \cite{karlsson2017future, JiaWenGeZhuWymTuf21}.

The bridge between communication and localization is the  propagation channel \cite{chen2022tutorial, JiaWenGeZhuWymTuf21, WenKulWitWym21}, since the communication quality is a function of the user equipment (UE), and the communication channel's geometric parameters (including the angle-of-arrival (AOA), and angle-of-departure (AOD)) and time-of-arrival (TOA)) are used both for communication and localization purposes \cite{WenKulWitWym21, JiaWenGeZhuWymTuf21, JiaGeZhuWym21}. 
Therefore, one of the main challenges in integrated localization and communication is to estimate the channel, from which one can localize the UE and optimize the communication quality. Following this idea, many channel estimation and localization studies have been conducted recently \cite{WenKulWitWym21, JiaWenGeZhuWymTuf21, JiaGeZhuWym21, ZhaHaa17, GeWenKimZhuJiaKimSveWym20}.
Generally,  channel estimation in massive MIMO mmWave communication systems can be divided into two categories: the \emph{on-grid} and the \emph{off-grid}  approaches. On-grid methods  rely on a pre-defined dictionary to estimate the multipath components \cite{Heath2016, lee2016channel}, and thus is inherently limited by the dictionary size. 
In contrast,  off-grid methods transform the channel estimation problem into an optimization problem \cite{tsai2018millimeter, liao2019closed, wang2019beam, wang2018spatial} or provide direct solutions based on the problem structure \cite{ZhaHaa17,SorLat16, LiuLiuMa07, SahUseCom17}. Among this last class of methods,  
estimation of signal parameters via rotational invariant techniques (ESPRIT) has been successful in mmWave massive MIMO channel estimation \cite{WenKulWitWym21, JiaWenGeZhuWymTuf21, JiaGeZhuWym21, liao2019closed, ZhaHaa17}. 

Most existing studies on ESPRIT  for mmWave massive MIMO channel estimation assume a narrowband channel model, which indicates that the bandwidth is relatively small and the array size is small \cite{WenKulWitWym21, JiaWenGeZhuWymTuf21, JiaGeZhuWym21, ZhaHaa17}. 
As pointed out in \cite{wang2019beam, wang2018spatial}, when the array size and the system bandwidth become large, 
the spatial wideband effects (also known as the \textit{beam-squint}) are non-negligible. In particular,  spatial wideband introduces an undesired coupling of the angular frequencies in the frequency domain (corresponding to the delays) and the spatial domain (corresponding to the angles).
To deal with the spatial wideband effects, the authors in \cite{wang2019beam, wang2018spatial} explore the optimization methods to find the angle and delay iteratively, which requires good initial estimates and high complexity due to the iterative search. In \cite{9049103} improved multidimensional folding (IMDF) per subcarrier was proposed in order to reconstruct the channel. However, the method requires several transmissions with varying channel gain and pairing of parameters among subcarriers and performance evaluation in terms of the spatial frequencies was not considered.

\begin{figure}
    \centering
    \includegraphics[width=0.8\linewidth]{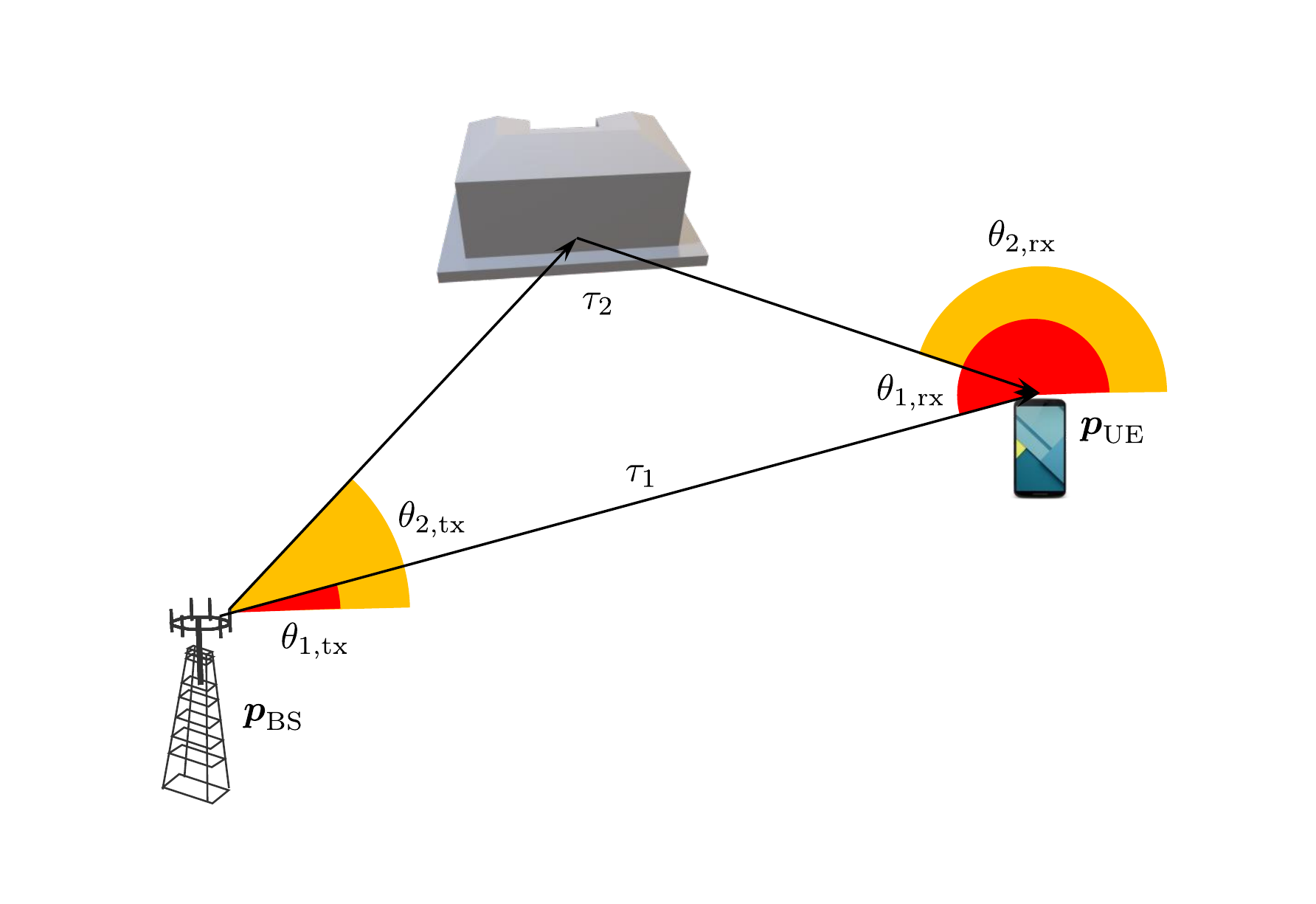} 
    \vspace{-4mm}
    \caption{Considered localization scenario with multipath. The objective of the UE is to estimate its location and clock bias based on the estimated AOA, AOD, and TOA of each path.}
    \label{fig:senario}
    \vspace{-5mm}
\end{figure}

In this paper, we investigate 
a low-complexity 
ESPRIT\footnote{As shown in \cite{SahUseCom17}, the MD-ESPRIT based approach outperforms the IMDF-based approach in terms of angular frequency estimations.} method for wideband mmWave massive MIMO channel estimation with explicit pairing across subcarriers. The proposed method can operate with a single snapshot and is evaluated in terms of both channel parameter estimation and positioning.

Our contributions are summarized as follows: (i) we propose a novel method for mmWave massive MIMO channel estimation under spatial wideband propagation based on a snapshot observation; (ii) we introduce a novel pairing strategy to deal with the auto-pairing problem of paths across subcarriers;  (iii) we provide a new narrowband condition for localization, and has verified the proposed condition in simulation; (iv) we assess the performance of the proposed method against standard approaches and show that the proposed leads to better performance, both in spatial frequency estimation and localization, even with relatively moderate spatial wideband effects.

\subsubsection*{Notations}

The operations $\otimes$ and $\odot$ denote the Kronecker and Khatri-Rao (column wise) products, respectively. The outer product is denoted by $\circ$, the Hadamard product (element wise) is denoted by $\bullet$, $\mathrm{vec_r}$ denotes vectorization by row\cite{SahUseCom17}, $(\cdot)$ denotes derivatives, and $\delta$ denotes differentiation operation.

\section{System Model}

Consider a massive MIMO orthogonal frequency division multiplexing (OFDM) system, which consists of a base station (BS) at location $\boldsymbol{p}_{\text{BS}}$ equipped with $M_{\text{tx}}$-antenna uniform linear array (ULA) and a UE  at location $\boldsymbol{p}_{\text{UE}}$ with $M_{\text{rx}}$-antenna ULA. Our objective is to estimate the UE location. 
There exist $L$ resolvable propagation paths between the BS and UE. The system bandwidth is $B$, and the number of subcarriers is $K$, i.e., the subcarrier spacing is $\Delta f = B/K$.  
The frequency domain channel, accounting for the spatial wideband effect \cite{wang2018spatial, wang2019beam}, over the $k$-th ($k \in \{ 0,  \cdots, K-1\}$) subcarrier from the $m_{\text{tx}}$-th ($m_{\text{tx}} \in \{ 0,  \cdots, M_{\text{tx}}-1\}$) BS antenna to the $m_{\text{rx}}$-th  ($m_{\text{rx}} \in \{ 0,  \cdots, M_{\text{rx}}-1\}$) UE antenna is given by 
\begin{align*}
& h_{m_{\text{tx}},m_{\text{rx}},k} =\\
& \sum_{l=1}^{L} \breve{\alpha}_{l} e^{-j2\pi m_{\text{tx}}\phi_{l,{\text{tx}},k}} e^{-j2\pi k\Delta f\tau_{l}}e^{-j2\pi m_{\text{rx}}\phi_{l,{\text{rx}},k}}, \label{eq1} \numberthis
\end{align*}
where for $i \in \{\text{tx},\text{rx}\}$,
\begin{align}
\phi_{l,k,i} & = \phi_{l,i}\big(1+\frac{k\Delta f}{f_{\text{c}}}\big),
\label{eq2}
\end{align}
$\breve{\alpha}_{l}=\alpha_{l}\mathrm{exp}({-j2\pi f_{\text{c}}\tau_{l}})$, where ${\alpha}_l$ denotes the complex gain of the $l$-th path, $\tau_{l}$ is the delay (TOA) of the $l$-th path, $\phi_{l,i}={d_i}\sin{\theta_{l,i}}/{\lambda_{\text{c}}}$  denotes the normalized AOD ($i=\text{tx}$) and AOA ($i=\text{rx}$), respectively, in which $\theta_{l,i}$ denotes the corresponding physical AOD and AOA of $l$-th path and $d_i$ denotes the antenna spacing at BS and UE, respectively. Finally, $f_{c}$ denotes the carrier frequency, and $\lambda_{c}$ is the corresponding wavelength.  For the line-of-sight (LOS) path, $\tau_{l}=\Vert\boldsymbol{p}_{\text{BS}}-\boldsymbol{p}_{\text{UE}} \Vert /c + \tau_B$,  where $\tau_B$ is the UE's clock bias, $\theta_{1,\text{tx}} = \mathrm{arctan}((p_{y,{\text{UE}}}-p_{y,{\text{BS}}})/(p_{x,{\text{UE}}}-p_{x,{\text{BS}}}))$, $\theta_{1,\text{rx}} = \mathrm{arctan}((p_{y,{\text{BS}}}-p_{y,{\text{UE}}})/(p_{x,{\text{UE}}}-p_{x,{\text{BS}}}))$. For the non-line-of-sight (NLOS) paths, supposed that the scatter of the $l$-th path locates at $\boldsymbol{p}_{l}$, $\tau_{l}=(\Vert\boldsymbol{p}_{\text{BS}}-\boldsymbol{p}_{l} \Vert+ \Vert\boldsymbol{p}_{\text{UE}}-\boldsymbol{p}_{l} \Vert)/c + \tau_B$, $\theta_{l,\text{tx}} = \mathrm{arctan}((p_{y,l}-p_{y,{\text{BS}}})/(p_{x,l}-p_{x,{\text{BS}}}))$, $\theta_{l,\text{rx}} = \mathrm{arctan}((p_{y,{\text{BS}}}-p_{y,l})/(p_{x,{\text{UE}}}-p_{x,l}))$.
 
We further assume that channel estimation is performed on pilot symbols or training sequence. Similar to \cite{JiaLiGon18_SPL}, we perform the maximum likelihood channel estimation, and the resultant channel estimate over the $k$-th subcarrier is given by
\begin{align}
\hat{\boldsymbol{H}}_k = \boldsymbol{H}_k + \delta \boldsymbol{H}_k, \label{eq3}
\end{align}
where $ \boldsymbol{H}_k \in \mathbb{C}^{M_{\text{tx}} \times M_{\text{rx}}}$ is constructed from \eqref{eq1}, $\delta \boldsymbol{H}_k$ denotes the channel estimation error with each entry modeled as zero mean independent and identically distributed Gaussian random variable. Finally, we assume $L$ is known. 

\section{Parallel ESPRIT-based Algorithm for Channel Parameter Estimation and Localization}
From \eqref{eq1}, we note that the AOAs, AODs, and delays are related to angular frequencies; as a result, the estimation of the channel parameters from noisy $\hat{\boldsymbol{H}}_k$ can be transformed into a multi-dimensional (MD) harmonic retrieval (MHR) problem. In this section, we formulate the corresponding tensor model, 
develop a parallel ESPRIT-based algorithm for channel parameter estimation and a pairing method across subcarriers. Based on the paired parameters, fusion across subcarriers is performed, after which the UE is  localized.
\subsection{Tensor Formulation of the Wideband Channel }\label{sec:channelTensor}
 
The channel model from \eqref{eq1} can be formulated with a 3D tensor $\boldsymbol{\mathcal{H}} \in C^{M_{\text{tx}} \times M_{\text{rx}} \times K}$, given by
\begin{equation}
\boldsymbol{\mathcal{H}}=\sum_{l=1}^{L}\breve{\alpha}_{l}\,\boldsymbol{a}^{(M_{\text{tx}})}_{l,1}\circ\boldsymbol{a}^{(M_{\text{rx}})}_{l,2}\circ\boldsymbol{a}^{(K)}_{l,3}\bullet \mathcal{\boldsymbol{D}}_l, \label{eq4} 
\end{equation}
where $\boldsymbol{a}^{(M_{\text{tx}})}_{l,1}$, $\boldsymbol{a}^{(M_{\text{rx}})}_{l,2}$, and $\boldsymbol{a}^{(K)}_{l,3}$ are steering vectors formed by $\mathrm{exp}({-j2\pi m_{\text{tx}}\phi_{l,\text{tx}}})$, $\mathrm{exp}({-j2\pi m_{\text{rx}}\phi_{l,\text{rx}}})$, and $\mathrm{exp}({-j2\pi k\Delta f\tau_{l}})$.

The phase rotation tensor $\mathcal{\boldsymbol{D}}_l \in C^{M_{\text{tx}} \times M_{\text{rx}} \times K}$ is produced by wideband effect, with entries $d_{l,m_{\text{tx}},m_{\text{rx}},k}=\mathrm{exp}({-j2\pi k\Delta f(m_{\text{tx}}\phi_{l,\text{tx}}/f_\text{c}+m_{\text{rx}}\phi_{l,\text{rx}}/f_\text{c}}))$. 
\begin{remark}[Narrowband model]
Note that without spatial wideband $\phi_{l,k,i}=\phi_{l,i}$, $\forall k$. Hence, all entries in $\mathcal{\boldsymbol{D}}_l$ tend to 1, so that standard 3D ESPRIT \cite{SahUseCom17} can be applied. 
\end{remark}

\begin{prop}[Revised narrowband condition]\label{NB Condition}
The narrowband model holds when $d_{l,m_{\text{tx}},m_{\text{rx}},k} \approx1$, which is fulfilled when  $2\pi(M_{\text{tx}}+M_{\text{rx}})K\Delta f d_i/({f_\text{c} \lambda_{\text{c}}})\ll 1$.
\end{prop}
\begin{proof}
The wideband effect impacts each path $l$ separately. Hence, we consider a single path $l$. 

All terms in ${\boldsymbol{D}}_l$ for the $l$-th path are of the form $d_{l,m_{\text{tx}},m_{\text{rx}},k}={\exp}({\phi_{l,m_{\text{tx}},m_{\text{rx}},k}})$. The  wideband effect causes a perturbation compared to narrowband case (where ${\exp}({\phi_{l,m_{\text{tx}},m_{\text{rx}},k}})=1$), hence makes the channel deviated from the narrowband case. 
The wideband effect causes a perturbation 
$
\delta\phi_{l,m_{\text{tx}},m_{\text{rx}},k}$ around $\phi_{l,m_{\text{tx}},m_{\text{rx}},k}=0$. 

We apply a perturbation analysis to derive the perturbation of the output when a small perturbation is added to the input \cite{1041039}, where in our case the input is $\phi_{l,m_{\text{tx}},m_{\text{rx}},k}$ and the output is ${{d}}_{l,m_{\text{tx}},m_{\text{rx}},k}$. 
 A Taylor expansion at ${\phi_{l,m_{\text{tx}},m_{\text{rx}},k}}=0$ yields (dropping all indices)
 \begin{align}
    & e^{\phi}  \approx  1+\left.\frac{\text{d}e^{\phi}}{ \text{d} \phi}\right|_{\phi=0}(\phi-0) = 1+ \phi. 
\end{align}
Thus the perturbation caused by input $\delta\phi_{l,m_{\text{tx}},m_{\text{rx}},k}$ leads to an identical output perturbation, i.e.,  $\delta d_{l,m_{\text{tx}},m_{\text{rx}},k}$. 

For the narrowband condition to hold, we require
\begin{align}
    \max_{m_{\text{tx}},m_{\text{rx}},k} |\delta d_{l,m_{\text{tx}},m_{\text{rx}},k}| \ll 1. \label{condition1}
\end{align}

The maximal possible values are given by $m_{\text{tx}}=M_{\text{tx}}-1$,  $m_{\text{rx}}=M_{\text{rx}}-1$, $k=K-1$, and $\phi_{l,i}={d_i}/{\lambda_{\text{c}}}$, which, combined with the form $d_{l,m_{\text{tx}},m_{\text{rx}},k}=\mathrm{exp}({-j2\pi k\Delta f(m_{\text{tx}}\phi_{l,\text{tx}}/f_\text{c}+m_{\text{rx}}\phi_{l,\text{rx}}/f_\text{c}}))$, leads to the specified condition.

\end{proof}
The new narrowband condition will be confirmed numerically in Section \ref{simulation}. For $d_i=\lambda_{\text{c}}/2 $, the condition could be further simplified to
$\pi(M_{\text{tx}}+M_{\text{rx}})B\ll f_\text{c}$.

\subsection{Proposed Method}
The proposed channel estimation method comprises 3 phases: per subcarrier processing, across subcarrier pairing, delay estimation. While a detailed complexity analysis is beyond the scope of the paper, we note that 
the complexity of the proposed method has the same scaling as \cite{9049103}. However, it  can perform channel estimation with only single transmission, while \cite{9049103} requires several transmissions with uncorrelated channels. 
\subsubsection{Phase 1: Parallel MD-ESPRIT}\label{sec:AlgorithmProp}
Extracting the channel response over the $k$-th subcarrier, we form the channel matrix $\boldsymbol{H}_k$, which can be written as
\begin{align}
\boldsymbol{H}_k=\sum_{l=1}^{L}\check{\alpha}_{l,k}\boldsymbol{a}^{(M_{\text{tx}})}_{l,k,\text{tx}}\boldsymbol{a}^{(M_{\text{rx}})\top}_{l,k,\text{rx}}, 
\label{eq5}
\end{align}
where $\check{\alpha}_{l,k}=\alpha_{l}\mathrm{exp}({-j2\pi (f_{\text{c}}+k\Delta f)\tau_{l}}$), and $\boldsymbol{a}^{(M_i)}_{l,k,i}$ is steering vector formed by modes of $l$-th path in $i$-th dimension, i.e., ${a}_{l,k,i}=\mathrm{exp}({-j2\pi(1+k\Delta f/f_{\text{c}})\phi_{l,i}})$.
By applying the MD-ESPRIT algorithm from \cite{SahUseCom17} to \eqref{eq4} for each $k$, the outputs are the paired modes $({a}_{l,k,\text{tx}},{a}_{l,k,\text{rx}})$. The paired angles $(\phi_{l,k,\text{tx}},\phi_{l,k,\text{rx}})$ associated with each subcarrier are then  calculated as
\begin{align}
\phi_{l,k,i}=-\frac{f_\mathrm{c}}{2\pi (f_\mathrm{c}+k\Delta f)}\mathrm{ln}({a}_{l,k,i}).
\end{align}
Although the angles are  auto-paired for each path for each subcarrier, the ordering is not maintained across subcarriers, i.e., $(\phi_{1,k,\text{tx}},\phi_{1,k,\text{rx}})$ and $(\phi_{1,k',\text{tx}},\phi_{1,k',\text{rx}})$ do not necessarily pertain to the same path $l$, for $k\neq k'$. This is because the singular values output of singular value decomposition (SVD) within MD-ESPRIT are in descending order, i.e., ordered based on received signal energy of the paths. Due to the noise, the order of signal energy of paths is not always  maintained across subcarriers. Therefore, all methods involving SVD techniques in a parallel architecture, such as \cite{9049103} and our parallel MD-ESPRIT, will encounter the pairing problem across subcarriers.

\subsubsection{Phase 2: Pairing Across Subcarriers}\label{sec:AlgorithmAngle}
To solve the pairing problem, one possible solution is to increase the SNR (by increasing the transmit power or pilot duration) so that noise is insignificant compared to signal energy. However, high transmit power is often not allowed due to hardware and energy limitations, while longer pilots reduce the effective data rate.
The traditional pairing method utilizing the same eigenstructure could not be applied here, since the eigenstructure of the channel on each subcarrier differ.
Our approach avoids the need for increasing the SNR and is based on a modified K-means algorithm \cite[Ch.~9]{bishop2006pattern}. 
We introduce $KL$ vectors of $\mathbf{y}_{(l-1)K+k}=[\phi_{l,k,\text{tx}},\phi_{l,k,\text{rx}}]^\top$. We now index these measurements as $\mathbf{y}_{n}$, $n=1,\ldots, N$, where $N=KL$. We randomly initialize K-means by choosing $L$ measurements. $\boldsymbol{\mu}_j\in [-\pi/2,\pi/2]\times [-\pi/2,\pi/2]$, $j=1,\ldots, L$. Then we iteratively update the assignments $r_{n,j}\in \{0,1\}$ and the means as 
\begin{align}
    r_{n,j}=
\begin{cases}
1\;\;\;\; j=\mathrm{argmin}_{j'} D(\mathbf{y}_{n}, \boldsymbol{\mu}_{j'})\\
0\;\;\;\;\mathrm{otherwise},
\end{cases}
\label{eq7}
\end{align}
and
\begin{align}
\boldsymbol{\mu}_j=\frac{ \sum_{n=1}^{N} r_{n,j} \mathbf{y}_n}{\sum_{n=1}^{N} r_{n,k}},
\label{eq8}
\end{align}
where $D(\cdot, \cdot)$ is a suitable distance metric that accounts for any angle wrapping. In our case, we have set $D(\mathbf{y}_{n}, \boldsymbol{\mu}_{j'})=\Vert \mathbf{y}_{n}-\boldsymbol{\mu}_{j'}\Vert$, as we operate in an SNR regime where angle wrapping is rare. 

We repeat \eqref{eq7}--\eqref{eq8} until there is no further change in the assignments, i.e., $r_{n,j}$ remains unchanged. Since the K-means algorithm is ignorant of the constraint that $[\phi_{l,k,\text{tx}},\phi_{l,k,\text{rx}}]^\top$ and $[\phi_{l',k,\text{tx}},\phi_{l',k,\text{rx}}]^\top$ are not allowed to be in the same cluster for $l' \neq l$, we check for each cluster if more than one measurement with the same subcarrier index are present. If so, we only retain the one closest to the cluster means and discard the others. Then the cluster means are recomputed. 
Finally, $\phi_{l,\text{tx}}$ and $\phi_{l,\text{rx}}$ are obtained from the means $[\phi_{l,\text{tx}},\phi_{l,\text{rx}}]^\top=\boldsymbol{\mu}_l$. Once $\phi_{l,i}$ is obtained, physical AOD and AOA are calculated by $\theta_{l,i}=\mathrm{arcsin}(\phi_{l,i}\lambda_{\text{c}}/d_i)$.

\begin{remark}
To reduce complexity, the K-means algorithm can be applied to only the AOA or AOD domain, provided the clusters are well separated in that domain. This is important when both elevation and azimuth angles are considered. 
\end{remark}

\subsubsection{Phase 3: Delay and Complex Gain Estimation}\label{sec:AlgorithmDelay}
It can be verified that the channel matrix over the $k$-th subcarrier admitted the following the form of matrix product
\begin{align}
\boldsymbol{{H}}_{k}=\boldsymbol{A}_{k,\text{tx}}^{(M_{\text{tx}})}\mathrm{diag}(\boldsymbol{\check{\alpha}}_{k})\boldsymbol{A}_{k,\text{rx}}^{(M_{\text{rx}})\top},
 \label{eq9}
\end{align}
where $\boldsymbol{\check{\alpha}}_{k}$ is vector containing $\check{\alpha}_{l,k}$ and $\boldsymbol{A}_{k,i}^{(M_i)}=[\boldsymbol{a}^{(M_i)}_{1,k,i}\cdots\boldsymbol{a}^{(M_i)}_{l,k,i}\cdots\boldsymbol{a}^{(M_i)}_{L,k,i}]$ can be reconstructed from the estimates of $\phi_{l,i}$ after pairing. The $\boldsymbol{\check{\alpha}}_{k}$ can further be calculated as
\begin{align}
    \boldsymbol{\check{\alpha}}_{k}=(\boldsymbol{A}_{k,\text{tx}}^{(M_{\text{tx}})}\odot\boldsymbol{A}_{k,\text{rx}}^{(M_{\text{rx}})})^\dagger\mathrm{vec_r}(\boldsymbol{{H}}_{k})
\end{align}
so that $\check{\alpha}_{l,k}$ can be obtained for all subcarriers by extracting diagonal elements of $\mathrm{diag}(\boldsymbol{\check{\alpha}}_{k})$. Take complex logarithm of $\check{\alpha}_{l,k}=\alpha_{l}e^{-j2\pi (f_{\text{c}}+k\Delta f)\tau_{l}}$ and stacking the values into a vector, we find that 
 \begin{align}
     \mathrm{ln}(\boldsymbol{\check{\alpha}}_{l})=\mathrm{ln}(\alpha_{l})\mathbf{1}+ \tau_{l}\boldsymbol{s}, 
       \label{peq11}
 \end{align}
 where ${s}_k=-j2\pi (f_{\text{c}}+k\Delta f)$. Introducing $\boldsymbol{S}=[\boldsymbol{s}\,\mathbf{1}]$, then the least squares (LS) estimate of delay and channel gain are 
 \begin{align}
 \begin{bmatrix}
     \tau_{l}\\
  \mathrm{ln}(\alpha_{l})
    \end{bmatrix}
  =\boldsymbol{S}^\dagger~\mathrm{ln}(\boldsymbol{\check{\alpha}}_{l}).
    \label{eq12}
 \end{align}
 For the noisy channel, we can take $\mathcal{R}\{ \tau_{l}\}$  as  the  estimated delay.

\subsection{Localization}\label{sec:AlgorithmLocal}
For line of sight (LOS), UE's location could be expressed as: (BS has synchronization error $ \tau_\mathrm{B}$)
\begin{align}
\boldsymbol{p}_{\text{UE}}=\boldsymbol{p}_{\text{BS}}+c(\tau_1- \tau_\mathrm{B})\boldsymbol{f}_{1,\text{rx}}
    \label{eq13}
\end{align}
For none light of sight (NLOS), UE's position can be expressed in $\theta_{l,\text{rx}}$, $\theta_{l,\text{tx}}$ and $\tau_l$ of each path according to geometry as \eqref{eq14} \cite{9179819} .
\begin{align}
\boldsymbol{p}_{\text{UE}}=\boldsymbol{p}_{\text{BS}}+\boldsymbol{f}_{l,\text{tx}}d_l+(c(\tau_l- \tau_\mathrm{B})-d_l)\boldsymbol{f}_{l,\text{rx}}
    \label{eq14}   
\end{align}
where $\boldsymbol{p}_{\text{UE}}$ and $\boldsymbol{p}_{\text{BS}}$ denote the UE's position and BS's position, respectively. $\boldsymbol{f}_{l,\text{tx}}$ and $\boldsymbol{f}_{l,\text{rx}}$ represent directional vectors corresponding to the departure and arrival of the signal, respectively, given by  $\boldsymbol{f}_{l,\text{rx}}=[\mathrm{cos}(\theta_{l,\text{rx}}), -\mathrm{sin}(\theta_{l,\text{rx}})]^\top$, $\boldsymbol{f}_{l,\text{tx}}=[\mathrm{cos}(\theta_{l,\text{tx}}), \mathrm{sin}(\theta_{l,\text{tx}})]^\top$. $d_l$ represents the travel distance of electromagnetic wave before reflection. Let $ \boldsymbol{f}_l=-(\boldsymbol{f}_{l,\text{rx}}+\boldsymbol{f}_{l,\text{tx}})$, joint equations given by \eqref{eq13} and \eqref{eq14} could be written as $\boldsymbol{B}\boldsymbol{v}=\boldsymbol{z}$, 
where
\begin{align}
\boldsymbol{B}=\begin{bmatrix}
    \boldsymbol{I}_{2\times2}&  c\boldsymbol{f}_{1,\text{rx}} & \boldsymbol{0}_{2\times1} & \boldsymbol{0}_{2\times1} &\cdots & \boldsymbol{0}_{2\times1}\\
    \boldsymbol{I}_{2\times2}&  c\boldsymbol{f}_{2,\text{rx}} & \boldsymbol{f}_2 & \boldsymbol{0}_{2\times1} & \cdots &\boldsymbol{0}_{2\times1}\\
    \boldsymbol{I}_{2\times2}&  c\boldsymbol{f}_{3,\text{rx}} & \boldsymbol{0}_{2\times1} & \ddots & \ddots & \vdots\\
    \vdots&  \vdots & \vdots & \ddots & \ddots & \boldsymbol{0}_{2\times1}\\
    \boldsymbol{I}_{2\times2}&  c\boldsymbol{f}_{L,\text{rx}} & \boldsymbol{0}_{2\times1} & \cdots & \boldsymbol{0}_{2\times1} & \boldsymbol{f}_{L}\\
\end{bmatrix}
\end{align}
and $\boldsymbol{v}=[
   \boldsymbol{p}_{\text{UE}}^\top \, \tau_\mathrm{B}\,  d_2 \cdots d_L
]^\top$, $\boldsymbol{z}=[
   \boldsymbol{p}_{\text{BS}}^\top + c\tau_l\boldsymbol{f}_{1,\text{rx}}^\top  \cdots \boldsymbol{p}_{\text{BS}}^\top + c\tau_L\boldsymbol{f}_{L,\text{rx}}^\top
]^\top.$

Using least squares, we have $\boldsymbol{v}=\boldsymbol{B}^\dagger\boldsymbol{z}$, the $\boldsymbol{p}_{\text{UE}}$ and $\tau_\mathrm{B}$  could be obtained by extracting the first three entries in $\boldsymbol{v}$. 
Measurements from total $L$ paths can establish $2L$ equations containing $L+2$ unknowns. The equations are over-determined when $2L\geq L+2$, and therefore can be solved when $L\geq 2$.

\section{Simulation}\label{simulation}

\subsection{System Setup}
The MIMO-OFDM localization scenario set up as follows: The BS is equipped with $M_{\text{tx}}=32$ ULA antennas at $(0,40)\mathrm{m}$, whereas the UE is equipped with $M_{\text{rx}}=32$ ULA antennas at $(40,0)\mathrm{m}$. The antenna spacing at both sides equals to half wavelength. The BS antenna send symbols with total transmitted power $P_\mathrm{t}=15$ dBm and synchronization error $\tau_\mathrm{B}=0\,\mathrm{s}$. The system bandwidth is $B=K\Delta f $ with subcarrier spacing $\Delta f=120\,\mathrm{kHz}$, the carrier frequency is $f_{\text{c}}=28\mathrm{GHz}$. 
For these parameter settings, the narrowband thresholds are 87.5 MHz \cite{wang2018spatial} and 13.9 MHz (from Proposition \ref{NB Condition}), where in both cases `$\ll$' is interpreted as `10 times smaller'. 
The number of pilots for obtaining the estimate \eqref{eq3} is  $n_\mathrm{p}=64$.
The (AOD, AOA) of paths are set as $(-45^\circ,45^\circ)$, $(-54^\circ,-62^\circ)$, $(65^\circ,40^\circ)$. The noise power spectral density (PSD) is $N_0=-174\,\mathrm{dBm(W/Hz)}$ and the noise figure (NF) $n_f=8\mathrm{dB}$. The energy loss of reflection for NLOS is set to $3\,\mathrm{dB}$.

We evaluate three different methods in terms of the RMSE of the channel parameters and the position: 
    (i) Standard 3D ESPRIT, which ignores the wideband effect\cite{SahUseCom17} and directly compute angles and delay; 
    (ii) The method from \cite{9049103} without auto-pairing, but replacing IMDF with MD-ESPRIT.  For angle estimation, we implement the proposed method without pairing step. For delay estimation described in Section \ref{sec:AlgorithmDelay}, we use $\phi_{l,k,i}$ to reconstruct $\boldsymbol{A}_{k,i}^{(M_i)}$; 
    (iii) The proposed method. 
For all methods, the localization is performed in the same way as Section \ref{sec:AlgorithmLocal}.
\subsection{Results and Discussion}
\begin{figure}
     \centering
     \begin{subfigure}[b]{0.45\textwidth}
         \centering
         \input{BLOSAOA} \vspace{0mm}
         \label{fig:angle}
     \end{subfigure}
     \hfill
     \begin{subfigure}[b]{0.45\textwidth}
         \centering
         \input{BLOSdelay} \vspace{0mm}
         \label{fig:delay}
     \end{subfigure}
     \hfill\vspace{-5mm}
      \caption{Channel parameter RMSE as a function of the bandwidth, for AOA estimation (top) and delay estimation (bottom), along with the narrowband (NB) conditions from \cite{wang2018spatial}  and Proposition \ref{NB Condition}.}
        \label{fig:para} \vspace{-5mm}
\end{figure}
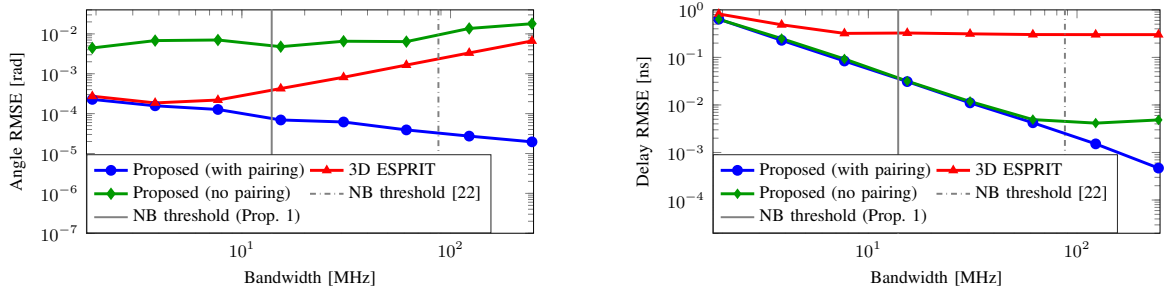

\subsubsection{Channel Parameter Estimation} 
In Fig.~\ref{fig:para}, we show the AOA\footnote{AOD performance is not shown as it is similar to AOA performance.} and delay estimation as a function of the bandwidth.  We observe 
that AOA estimation RMSE from 3D ESPRIT first improves with increasing bandwidth, as the overall SNR increases. With further increases in bandwidth, however, the wideband effect becomes dominant, and the RMSE quickly degrades. 
In contrast, the AOA RMSE of proposed method always improves with more bandwidth. The importance of correct pairing across subcarriers is evident from the high of the proposed method without pairing. 
In terms of  delay RMSE,  the 3D ESPRIT method exhibits near-constant performance.  This is because 3D ESPRIT doing averaging in each dimension, and the dimension containing delay information is not affected by the wideband effect. For the proposed method, the delay RMSE keeps decreasing due to improving accuracy of $\boldsymbol{A}_{k,i}^{(M_i)}$ reconstruction in \eqref{eq9} as a result of improved angle estimation. The delay estimation without pairing first improves due to increasing SNR, then remains stable, as it is limited by reconstruction accuracy of $\boldsymbol{A}_{k,i}^{(M_i)}$, since the angle estimation per subcarrier are of the same accuracy and there is no way to harness SNR gain by fusing across the subcarriers.
\subsubsection{Localization}
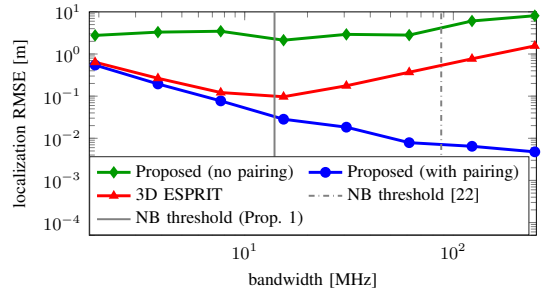
\begin{figure}[!htb]
    \centering
    \scalebox{.45}{\input{BLOSlocalization.tex} }
    \caption{Localization RMSE as a function of the bandwidth for the proposed method (with and without pairing), compared to the standard 3D ESPRIT with auto-pairing, along with the narrowband (NB) conditions from \cite{wang2018spatial}  and Proposition \ref{NB Condition}.} \vspace{-5mm}
    \label{fig:local-nc}
\end{figure}

From Fig.~\ref{fig:local-nc}, we observe that the localization RMSE of the proposed method monotonically improves when larger bandwidth is applied as a consequence of continuously improving accuracy of angle and delay estimation. The localization RMSE of wideband 3D ESPRIT initially goes down due to improvement of delay estimation, however, it starts to diverge after the about 15 MHz due to poor angle RMSE. The localization without proper pairing causes large error, due to the poor AOA and AOD estimates.  
From Figs.~\ref{fig:para}--\ref{fig:local-nc}, we see that the wideband effect has taken place before the  threshold in \cite{wang2018spatial}.

\section{Conclusions}

In this paper, we proposed a new method to  deal with the spatial wideband effect with snapshot observations. As part of this method, we investigate the pairing problem that appears in  methods based on per-subcarrier AOA and AOD estimation, and proposed a strategy to address this problem. Furthermore, we provided a tighter narrowband condition. Finally, we demonstrate that the wideband effect significantly impacts localization performance, if not properly addressed.

\balance

\end{document}

%% file: BLOSAOA.tex
\begin{tikzpicture}[scale=1\columnwidth/10cm,font=\footnotesize]
\begin{axis}[%
width=8cm,
height=4cm,
scale only axis,
xmin=1.8,
xmax=250,
ymode=log,
xmode=log,
ymin=0.0000001,
ymax=0.04,
xlabel={Bandwidth [MHz]},
ylabel={Angle RMSE [rad]},
yminorticks=true,
xminorticks=true,
axis background/.style={fill=white},
legend style={at={(0,0)}, anchor=south west, legend columns=2, legend cell align=left, align=left, draw=white!15!black}
]

\addplot [color=blue, line width=1.5pt, mark=*]
  table[row sep=crcr]{%
  1.9200  0.000227505431116485\\
 3.8400   0.000157296107171407\\
  7.6800	0.000127102775587922\\
15.3600  6.93498056221184e-05\\
30.7200	 6.17396200132643e-05\\
61.4400	 3.90692433594859e-05\\
122.8800	2.72960160358160e-05\\
245.7600	1.96035771049097e-05\\
};
\addlegendentry{Proposed (with pairing)}



\addplot [color=red, line width=1.5pt, mark size=1.5pt, mark=triangle*, mark options={solid, fill=red, draw=red}]
  table[row sep=crcr]{%
    1.9200  0.000273331471129645\\
 3.8400   0.000185713488424962\\
7.6800		0.000219393131129027\\
15.3600  	0.000425728896671505\\
30.7200		0.000817024947802086\\
61.4400		0.00165275362434702\\
122.8800	0.00332517241511647\\
245.7600	0.00664988234703270\\
};
\addlegendentry{3D ESPRIT}



\addplot [color=OliveGreen, line width=1.5pt, mark size=2pt, mark=diamond*, mark options={solid, fill=OliveGreen, draw=OliveGreen}]
  table[row sep=crcr]{%
 1.9200  0.004434992582607\\
 3.8400   0.00678665443200\\
 7.6800	0.00702632499258220\\
15.3600 0.00477512863092607\\
30.7200	0.00654099508572445\\
61.4400	0.00639748231144193\\
122.8800	0.0136857387335452\\
245.7600	0.0179836622235350\\
};
\addlegendentry{Proposed (no pairing)}


\addplot [color=black!50, dashdotted, line width=1pt]
  table[row sep=crcr]{
87.5	0.0000001\\
87.5	100\\
};
\addlegendentry{NB threshold \cite{wang2018spatial}}

\addplot [color=black!50,   line width=1pt]
  table[row sep=crcr]{
13.9	0.0000001\\
13.9	100\\
};
\addlegendentry{NB threshold (Prop.~1)}

\end{axis}

\end{tikzpicture}%

%% file: BLOSdelay.tex
\begin{tikzpicture}[scale=1\columnwidth/10cm,font=\footnotesize]
\begin{axis}[%
width=8cm,
height=4cm,
scale only axis,
xmin=1.8,
xmax=250,
ymode=log,
xmode=log,
ymin=0.00002,
ymax=1,
xlabel={Bandwidth [MHz]},
ylabel={Delay RMSE [ns]},
yminorticks=true,
xminorticks=true,
axis background/.style={fill=white},
legend style={at={(0,0)}, anchor=south west, legend columns=2, legend cell align=left, align=left, draw=white!15!black}
]

\addplot [color=blue, line width=1.5pt, mark=*]
  table[row sep=crcr]{%
1.9200  0.637553778128083\\
3.8400   0.229100987777723\\
7.6800	0.0840518164091322\\
15.3600 0.0310018670126833\\
30.7200	0.0110628583779995\\
61.4400	0.00422313108673359\\
122.8800	0.00151752423040202\\
245.7600	0.000469736033717492\\
};
\addlegendentry{Proposed (with pairing)}

\addplot [color=red, line width=1.5pt, mark size=1.5pt, mark=triangle*, mark options={solid, fill=red, draw=red}]
  table[row sep=crcr]{%
 1.9200  0.820703618276360\\
3.8400    0.486168397585780\\
 7.6800	0.320301865030639\\
15.3600 0.326193518046247\\
30.7200	0.313341706762633\\
61.4400	0.302666375696563\\
122.8800	0.301593968171360\\
245.7600	0.301670160118593\\
};
\addlegendentry{3D ESPRIT}



\addplot [color=OliveGreen,line width=1.5pt, mark size=1.5pt, mark=diamond*, mark options={solid, fill=OliveGreen, draw=OliveGreen}]
  table[row sep=crcr]{%
 1.9200 0.63491808128083\\
 3.8400  0.24917777009872\\
 7.6800	0.0927675599651816\\
15.3600 0.0317460220729517\\
30.7200	0.0118349184726598\\
61.4400	0.00491807316843619\\
122.8800	0.00414769240267890\\
245.7600	0.00483996161040421\\
};

\addlegendentry{Proposed (no pairing)}

\addplot [color=black!50, dashdotted,  line width=1pt]
  table[row sep=crcr]{
87.5	0.00001\\
87.5	100\\
};
\addlegendentry{NB threshold \cite{wang2018spatial}}

\addplot [color=black!50,   line width=1pt]
  table[row sep=crcr]{
13.9	0.00001\\
13.9	100\\
};
\addlegendentry{NB threshold (Prop.~1)}

\end{axis}

\end{tikzpicture}%

%% file: BLOSlocalization.tex
\begin{tikzpicture}[scale=1\columnwidth/10cm,font=\footnotesize]
\begin{axis}[%
width=8cm,
height=4cm,
scale only axis,
xmin=1.8,
xmax=250,
ymode=log,
xmode=log,
ymin=0.00005,
ymax=10,
xlabel={bandwidth [MHz]},
ylabel={localization RMSE [m]},
yminorticks=true,
xminorticks=true,
axis background/.style={fill=white},
legend style={at={(0,0)}, anchor=south west, legend columns=2, legend cell align=left, align=left, draw=white!15!black}
]


\addplot [color=OliveGreen, line width=1.5pt, mark size=2pt, mark=diamond*, mark options={solid, fill=OliveGreen, draw=OliveGreen}]
  table[row sep=crcr]{%
1.9200 2.76555890544330\\
3.8400 3.30653789435711\\
7.6800	3.47118610029134\\
15.3600	2.12796676405171\\
30.7200	2.92627340072738\\
61.4400	2.81289279417556\\
122.8800	6.07429759748241\\
245.7600	8.10917831312765\\
};
\addlegendentry{Proposed (no pairing)}

\addplot [color=blue, line width=1.5pt, mark=*]
  table[row sep=crcr]{%
 1.9200 0.546396302739816\\
 3.8400 0.194814130435810\\
7.6800	0.0771720477195207\\
15.3600	0.0283397188939211\\
30.7200	0.0183542019527428\\
61.4400	0.00787358329150054\\
122.8800	0.00645864941987988\\
245.7600	0.00476778031746880\\
};
\addlegendentry{Proposed (with pairing)}

\addplot[color=red, line width=1.5pt, mark size=1.5pt, mark=triangle*, mark options={solid, fill=red, draw=red}]
  table[row sep=crcr]
 {
 1.9200 0.638314149845030\\
 3.8400 0.265409468657441\\
7.6800	0.121727940584342\\
15.3600	0.0968547958439508\\
30.7200	0.176265693495103\\
61.4400	0.369300808470257\\
122.8800	0.768689100182696\\
245.7600	1.55791019505083\\
};
\addlegendentry{3D ESPRIT}

\addplot [color=black!50,  dashdotted, line width=1pt]
  table[row sep=crcr]{
87.5	0.0001\\
87.5	100\\
};
\addlegendentry{NB threshold \cite{wang2018spatial}}

\addplot [color=black!50, line width=1pt]
  table[row sep=crcr]{
13.9	0.0001\\
13.9	100\\
};
\addlegendentry{NB threshold (Prop.~1)}


\end{axis}
\end{tikzpicture}